\documentclass[pra,aps,twocolumn,nopacs,superscriptaddress,nofootinbib]{revtex4}
\usepackage{amsmath}  \usepackage{amssymb}  \usepackage{amsfonts}  \usepackage{bm}  \usepackage{bbm}   \usepackage{bbold}   \usepackage{braket}  \usepackage{color}  \usepackage{comment}  \usepackage{dcolumn}  \usepackage{enumerate}  \usepackage{epsfig}  \usepackage{gensymb}  \usepackage{graphicx}  \usepackage{indentfirst}  \usepackage{lmodern}  \usepackage{mathrsfs}  \usepackage{mathtools}  \usepackage{psfrag}  \usepackage{pst-all}  \usepackage{soul}  \usepackage{units}  \usepackage{xcolor}
\usepackage{float} %---APS---SI---arxiv
\usepackage[colorlinks,linkcolor=blue,citecolor=blue,urlcolor=blue,hyperindex,driverfallback=dvipdfm]{hyperref}  \usepackage[T1]{fontenc} %---APS---ACS---SI---arxiv

\def\ii{{\rm i}}  \def\ee{{\rm e}}
\def\me{m_{\rm e}}  
  
        \def\Eb{{\bf E}}                                \def\Qb{{\bf Q}}    \def\Rb{{\bf R}}  \def\rb{{\bf r}}      \def\vb{{\bf v}} %--- bold vectors
\def\xx{\hat{\bf x}}    \def\zz{\hat{\bf z}}        \def\eh{\hat{\bf e}}    
\def\blue{\color{blue}}  \def\red{\color{red}}  \def\orange{\color{orange}}    
\def\Rmax{{R_{\rm max}}}    \def\vv{\hat{\bf v}}

\begin{document} %---APS---SI---arxiv

\title{Spatiotemporal Electron-Beam Focusing through Parallel Interactions
\\ with Shaped Optical Fields}

\author{F.~Javier~Garc\'{\i}a~de~Abajo}
\email{javier.garciadeabajo@nanophotonics.es} %---
\affiliation{ICFO-Institut de Ciencies Fotoniques, The Barcelona Institute of Science and Technology, 08860 Castelldefels (Barcelona), Spain} %---
\affiliation{ICREA-Instituci\'o Catalana de Recerca i Estudis Avan\c{c}ats, Passeig Llu\'{\i}s Companys 23, 08010 Barcelona, Spain} %---

\author{Claus~Ropers}
\affiliation{Max Planck Institute for Multidisciplinary Sciences, 37077 G\"ottingen, Germany} %---
\affiliation{4th Physical Institute – Solids and Nanostructures, University of G\"ottingen, 37077 G\"ottingen, Germany} %---

\begin{abstract}
The ability to modulate free electrons with light has emerged as a powerful tool to produce attosecond electron wavepackets. However, research has so far aimed at the manipulation of the longitudinal wave function component, while the transverse degrees of freedom have primarily been utilized for spatial rather than temporal shaping. Here, we show that the coherent superposition of parallel light-electron interactions in separate spatial zones allows for a simultaneous spatial and temporal compression of a convergent electron wave function, enabling the formation of sub-{\AA}ngstr\"om focal spots of attosecond duration. The proposed approach will facilitate the exploration of previously inaccessible ultrafast atomic-scale phenomena in particular enabling attosecond scanning transmission electron microscopy.
\end{abstract}

\maketitle %---APS---OSA---SI---arxiv
\date{\today} %---APS---arxiv

% =========================================================
% --- collimated e-beam interaction with lossy sample -----
% =========================================================
%\section{...} \label{sec1}

% =========================================================
\section{Introduction}

Fourier analysis shows that strongly peaked waveforms can be obtained by superimposing a large number of phase-locked frequency components. This ubiquitous principle underpins pulsed mode-locked lasers and is leveraged to synthesize attosecond light pulses by combining high harmonics generated in atomic gases \cite{PTB01,CK07,KI09} or solid-state targets \cite{LLC20}. Likewise, attosecond electron bunches were formed through interaction with the near fields induced by laser scattering at a periodic structure followed by electron propagation in free space \cite{SCI08}. In addition, optical near-field interaction and dispersive propagation were predicted to produce attosecond wavepackets in the wave function of individual electrons \cite{FES15}, as later demonstrated in experiments using laser scattering by nanostructures \cite{PRY17,MB18_2} and also through stimulated Compton scattering in free space \cite{KSH18}.

Temporal compression of free-electron beams (e-beams) has a long tradition in the context of accelerator physics and electromagnetic wave generation \cite{SCC05,SCI08,G11_2,AB04,EAA10,HSX14,PMR16,GIF19,RTN20}. In an intuitive picture, exposure of the e-beam to electromagnetic fields induces a momentum modulation that causes a periodic compression into subcycle bunches upon dispersive propagation of the electron ensemble. By subsequently interacting with gratings \cite{SP1953} or undulators \cite{PMR16}, bunches containing a large number of electrons $N$ can produce radiation by acting in unison, giving rise to directed emission with an intensity $\propto N^2$ in what is known as superradiance \cite{UGK98}. This mechanism is widely used in radiation sources operating over spectral ranges extending from microwaves in klystrons \cite{G11_2} to x-rays in free-electron lasers \cite{AB04,EAA10,PMR16,GIF19}.

% Figure 1 -----------------------------------------------
\begin{figure*}
\begin{centering} \includegraphics[width=1.00\textwidth]{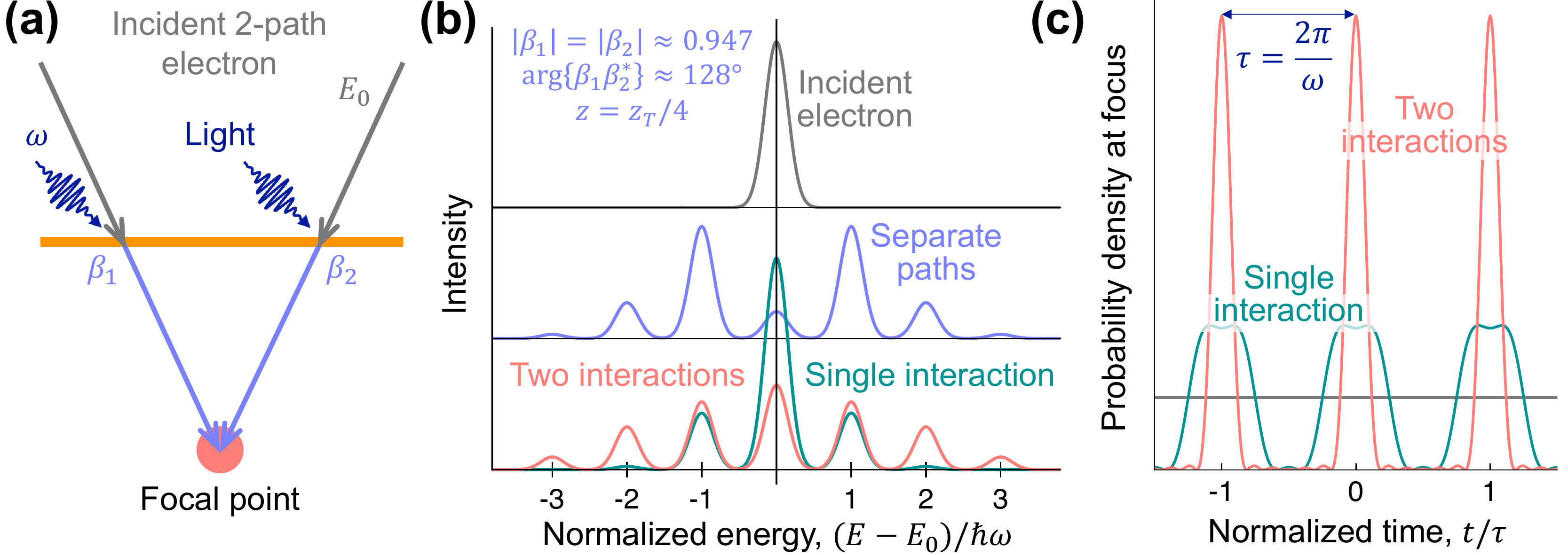} \par\end{centering}
\caption{{\bf Temporal compression of a free electron through parallel interactions with light.} (a) We consider an incident electron separated into two paths, each of them interacting with light as described by the respective coupling coefficients $\beta_1$ and $\beta_2$. The two paths are then recombined at a focal point. (b) Spectrum of the incident electron state before (top) and right after (middle) interacting with light, along with the spectrum produced at the focal point by path mixing (bottom). (c) Temporal profile at the focal point, showing a comb of electron probability density with the same period $\tau=2\pi/\omega$ as the employed light. The interaction coefficients in (b) and (c) are optimized to achieve maximum temporal compression. We also show results for an optimally compressed single-path/interaction configuration.}
\label{Fig1}
\end{figure*}

% Figure 2 -----------------------------------------------
\begin{figure}
\begin{centering} \includegraphics[width=0.45\textwidth]{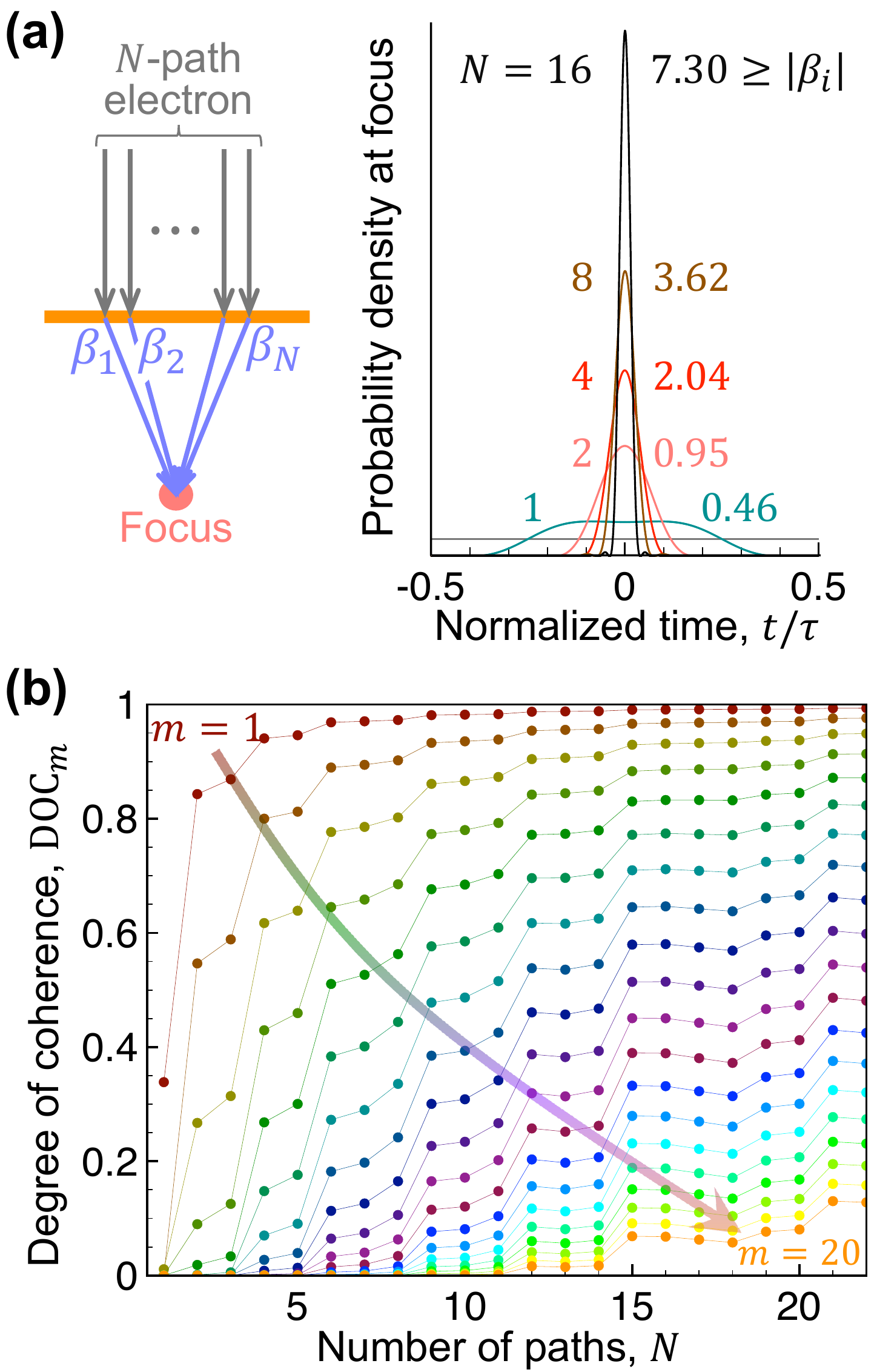} \par\end{centering}
\caption{{\bf Optimum temporal compression.} (a) An increasing number $N$ of parallel interactions with light leads to a narrowing of the probability density peak. The maximum magnitude of the required coupling coefficient is indicated for each $N$ by color-matching labels. Optimum results are obtained for a propagation distance $z=z_T/4$ in all cases. (b) Degree of coherence ${\rm DOC}_m$ for optimally compressed electrons after interaction with $N$ parallel zones. We consider harmonic orders $m=1-22$.}
\label{Fig2}
\end{figure}

% Figure 3 -----------------------------------------------
\begin{figure*}
\begin{centering} \includegraphics[width=0.65\textwidth]{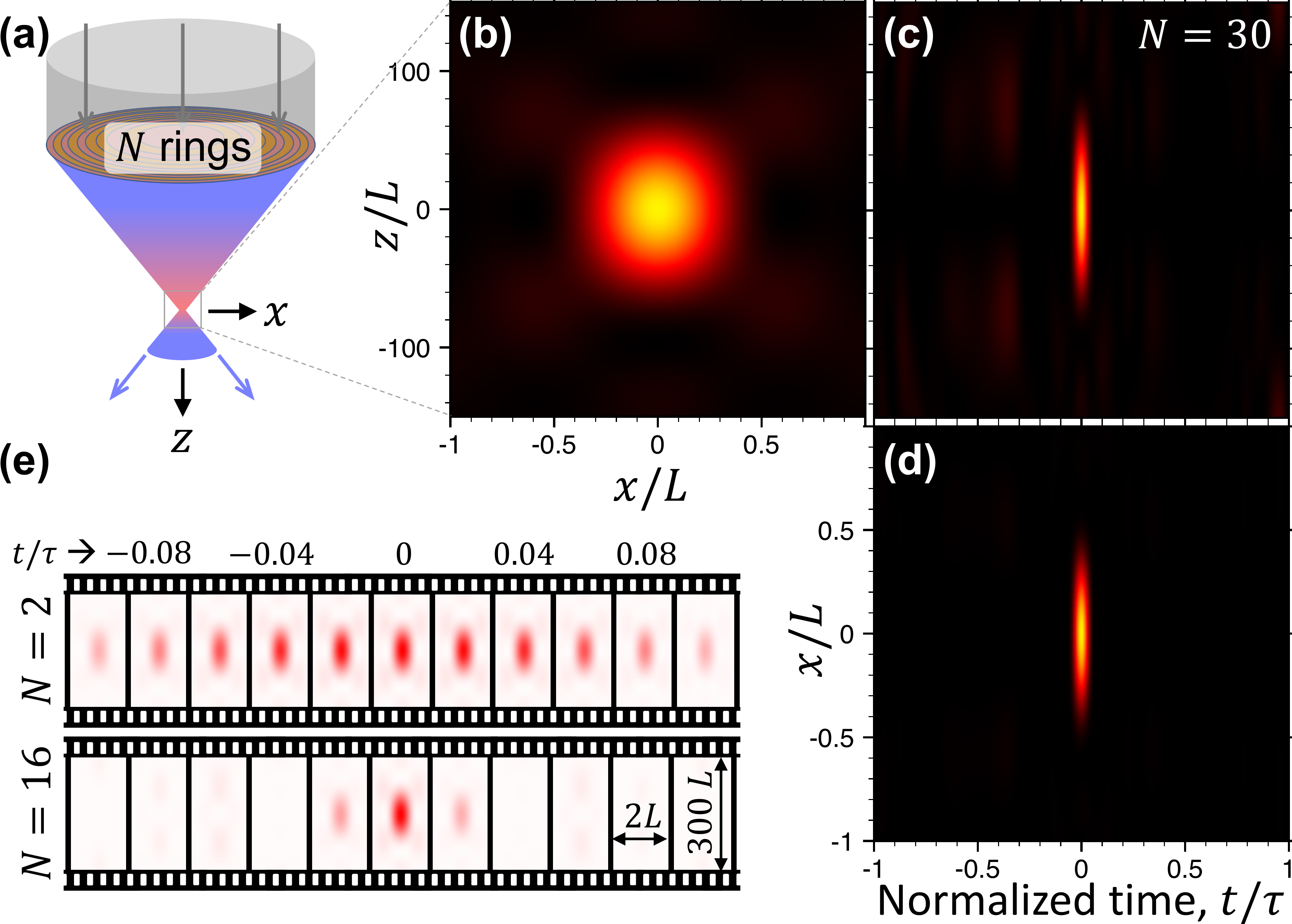} \par\end{centering}
\caption{{\bf Focal spot profile under temporal compression.} (a) We consider a set of $N$ equal-area concentric circular zones, each of them delivering a uniform light-electron coupling coefficient. (b-d) Upon optimization of the latter for electron temporal compression at the focal spot $\rb=0$ using $N=30$ zones, we still find a Gaussian-like spatial profile (b) of short duration, as revealed in cuts along both longitudinal (c) and transverse (d) directions. (e) A selection of focal region frames near the time of maximum electron probability reveals a short duration compared with the optical period $\tau$. The pulse duration is shown to decrease with increasing $N$. All distances are scaled to $L=\lambda_e/{\rm NA}$, where $\lambda_e$ is the de Broglie electron wavelength and NA is the numerical aperture (outer-zone diameter divided by zone-focus distance).}
\label{Fig3}
\end{figure*}

Electron compression can also be accomplished through the coherent evolution of each individual free-electron wave function after interaction with intense optical fields, provided that the level of spatial and temporal coherence of both electrons and light is sufficiently high.
% In particular, the evanescent fields produced upon light scattering by a structure can exchange real photons with the electron with a probability proportional to the strength of the optical response of the material. This process, which was proposed to realize electron energy-gain spectroscopy (EEGS) \cite{paper114} combining the spatial precision of e-beams and the spectral sharpness of laser sources, has been recently demonstrated with simultaneous nanometer and microelectronvolt resolution \cite{HRF21,FKS22}. 
For sufficient laser intensities, multiple photon exchanges take place, in analogy to low-energy electron scattering by illuminated atomic targets \cite{WHC1977,WHS1983}. In the context of ultrafast transmission electron microscopy, this type of process has attracted considerable interest in the form of photon-induced near-field electron microscopy (PINEM) of optical near-field distributions \cite{BFZ09,paper151,FES15,PLQ15,KLS20,WDS20,T20}. In PINEM, electrons emerge in states comprising a superposition of energy sidebands equally spaced by the photon energy. Besides imaging, the quantum-coherent phase modulation underlying the inelastic light-electron scattering process was predicted \cite{FES15} and experimentally shown \cite{PRY17,MB18_2} to produce longitudinal shaping and attosecond bunching. In the momentum representation, the velocity dispersion across the sidebands translates into relative phase differences that accumulate as the probes propagate, developing a periodic train of temporally compressed electron pulses analogous to the Talbot effect \cite{PRY17,paper360,TRB21}. % that can be reconstructed by quantum state tomography \cite{PRY17}.

While a single PINEM interaction is fundamentally limited to produce just a moderate level of temporal compression \cite{ZSF21,paper374,paper373}, close to perfectly confined pulses are predicted to be formed from a sequence of PINEM interactions separated by free-space propagation \cite{YFR21}. In a separate development, following the demonstration of ponderomotive phase plates for electron microscopy \cite{SAC19}, the possibility of realizing a customizable modulation of the transverse electron wave function profile was proposed using PINEM \cite{paper351} and ponderomotive \cite{paper368} light-electron interactions, and recently realized in separate experiments \cite{paper397,MWS22}. Conceivably, the coherent superposition of electron waves that have undergone distinct PINEM interactions in the transverse plane should grant us access into a much wider range of electron wave functions such as, for example, states that comprise tailored spatiotemporal compression.

% \cite{paper306,WDS20}

In this work, we theoretically demonstrate that inelastic electron--light interaction can simultaneously produce spatial and temporal compression. Specifically, we consider a convergent electron wave produced by the objective lens of a scanning transmission electron microscope and study the effect of the interaction with light at a plane preceding the focal plane. For relatively simple transverse field profiles reminiscent of zone plates, we predict the formation of sub-{\AA}ngstr\"om focal spots of attosecond duration. In particular, a high level of compression is achieved with the superposition of only two wave function components (Fig.~\ref{Fig1}), while more complex profiles (Figs.~\ref{Fig2} and \ref{Fig3}) enable a stronger temporal compression without substantially compromising the spatial focusing performance of the microscope. Our work holds potential for the study of ultrafast phenomena at the atomic scale, including highly nonlinear and subcycle charge and lattice dynamics.

% =========================================================
\section{Results and discussion}

Right after interaction with monochromatic light of frequency $\omega$ characterized by a space- and time-dependent electric field amplitude $\Eb(\rb)\ee^{-\ii\omega t}+{\rm c.c.}$, the wave function of an electron moving along the $z$ direction with average velocity $v$ becomes \cite{paper151,PLZ10,paper371}
\begin{align}
\psi(\rb,t)=\psi^{\rm inc}(\rb,t)\sum_{\ell=-\infty}^\infty \alpha_\ell(\rb) \ee^{\ii\ell \omega(z-vt)/v},
\label{mpsi1}
\end{align}
where $\psi^{\rm inc}(\rb,t)$ is the incident wave function. The sum in Eq.~(\ref{mpsi1}) extends over the net number of exchanged photons $\ell$, corresponding to an electron energy change $\ell\hbar\omega$ and having an associated transition amplitude
\begin{align}
\alpha_\ell(\rb)=J_\ell\big[|\beta(\Rb)|\big]\,\ee^{\ii\ell\,{\rm arg}\{-\beta(\Rb)\}}\,\ee^{-2\pi\ii\ell^2 z/z_T},
\label{malpha1}
\end{align}
which is expressed in terms of a single electron-light coupling parameter
\begin{align}
\beta(\Rb)=\frac{e}{\hbar\omega}\int_{-\infty}^\infty dz\, E_z(\rb)\,\ee^{-\ii\omega z/v}.
\label{mbeta}
\end{align}
We note that Eq.~(\ref{mbeta}) depends on the transverse coordinates $\Rb=(x,y)$. This result assumes an initial energy spread much smaller than $\hbar\omega$, which is in turn negligible compared with the average kinetic energy (nonrecoil approximation). In addition, a phase $\propto\ell^2$ is incorporated to include the effect of velocity dispersion with a characteristic Talbot distance \cite{paper360} $z_T=4\pi\me v^3\gamma^3/\hbar\omega^2$, where $\gamma=1/\sqrt{1-v^2/c^2}$.

In the spatial representation, the wave function is modulated in time with the optical period $\tau=2\pi\omega/v$ imposed by the light frequency $\omega$. This allows for a quantification of the achieved level of temporal compression through the so-called degree of coherence \cite{paper374,paper373}
\begin{align}
{\rm DOC}_m(\rb)=\bigg|\!\!\int_0^\tau\!\!\!\!dt\,|\psi(\rb,t)|^2\,\ee^{\ii m\omega t}\bigg|^2\!\!\bigg/\bigg|\!\!\int_0^\tau\!\!\!\!dt\,|\psi(\rb,t)|^2\bigg|^2\!\!, \label{mDOC}
\end{align}
which measures the ability of the electron to excite an optical mode localized at a position $\rb$ and characterized by a harmonic frequency $m\omega$. The ideal compression associated with the point-particle limit is thus corresponding to ${\rm DOC}_m(\rb)=1$ for all $m$'s.

For a single light-electron interaction, one finds the analytical expression \cite{ZSF21,paper373,YFR21,TRB21} ${\rm DOC}_m =J_m^2[4\beta\sin(2\pi mz/z_T)]$. For $m=1$, a moderate maximum value of ${\rm DOC}_1\approx0.34$ [see Fig.~\ref{Fig1}(b)] is obtained by adjusting the optical field strength to satisfy the condition $|\beta|=e_1/4\sin(2\pi z/z_T)$, where $e_1\approx1.841$ is the absolute maximum of the $J_1$ Bessel function. As stated above, even tighter compression can be reached by sequential interactions \cite{YFR21}. However, harnessing the transverse degrees of freedom to produce tailored temporal as well as spatial structuring has yet to be explored.

Here, we leverage the coherent superposition of electron wave function components undergoing separate parallel interactions with light in distinct zones for far-reaching spatiotemporal control. Replacing the coefficients in Eq.~(\ref{malpha1}) by a weighted sum with various values of $\beta$ yields a powerful set of additional control parameters. In a simple picture, temporal compression can be achieved if $\alpha_\ell$ becomes independent of $\ell$ over a certain range such as $-L\le\ell\le L$, for which the wave function becomes $\psi\propto\sum_{\ell=-L}^L\ee^{\ii\ell\zeta}=\sin[(L+1/2)\zeta]/\sin(\zeta/2)$, such that ${\rm DOC}_m\approx(1-|m|/L)^2$ approaches the perfect compression limit for $L\gg1$. The question we ask is whether near-unity, $\ell$-independent coefficients can be obtained by superimposing parallel electron-light interactions, such that they become
\begin{align}
\alpha_\ell(\rb)=\ee^{-2\pi\ii\,\ell^2z/z_T}\sum_i a_i \,J_\ell(2|\beta_i|) \,\ee^{\ii\ell{\rm arg}\{-\beta_i\}} \label{malpha2}
\end{align}
for a given set of weighting coefficients $a_i$, with a common dispersive phase $-2\pi z/z_T$ proportional to the distance $z$ from a given interaction plane to a common focal spot toward which the electron is converged.

The superposition of wave function components from {\it two} parallel zones [Fig.~\ref{Fig1}(a)] is already enough to produce a substantial improvement in temporal compression corresponding to ${\rm DOC}_1\approx0.84$ \cite{notecompression1} and illustrated by the sharp wave function profile plotted in Fig.~\ref{Fig1}(c), where it is compared with the single-interaction result. The spectral distribution of the wave function resulting from this superposition cannot be achieved with a single interaction [Fig.~\ref{Fig1}(b)]; it rather approaches a more continuous spectral distribution, as also observed for sequential interactions \cite{YFR21}. This exemplifies a method to produce any designated combination of sideband amplitudes by superimposing a sufficient number of parallel interactions. Incidentally, we impose real coefficients $a_i$ in Fig.~\ref{Fig1} (see below), for which the optimum solution involves moderate values of the coupling coefficients $\beta_i$ [see Fig.~\ref{Fig1}(b)].

As a practical zone-plate-type configuration, we consider an $\Rb$-dependent light-electron interaction taking place at a plane situated within the pole piece gap in an electron microscope and before the focal plane, although similar designs could operate by placing the plate at other places along the electron column. Under the paraxial approximation, and assuming axially symmetric coupling coefficients $\beta(\Rb)$ with respect to the e-beam axis at $\Rb=0$, the wave function near the focal region is found to take the form of Eq.~(\ref{mpsi1}) with coefficients (see details in the Appendix)
\begin{align}
\alpha_\ell(\rb)\approx\ee^{-2\pi\ii\,\ell^2z/z_T}&\int_0^{\rm NA} \!\!\! \theta d\theta \,J_0(2\pi\theta R/\lambda_e) \,\ee^{-\ii\pi\theta^2 z/\lambda_e} \nonumber\\
&\times J_\ell\big[2|\beta(\theta)|\big] \,\ee^{\ii\ell{\rm arg}\{-\beta(\theta)\}}, \label{malpha3}
\end{align}
where NA is the numerical aperture (set to 0.02 in this work), and $\lambda_e$ is the electron wavelength. This expression is accurate within the paraxial approximation for the wide range of geometrical parameters encountered in currently available electron microscopes (see Appendix). Incidentally,
both spherical and chromatic aberration can easily be included in our formalism, but are not considered here for simplicity. For concreteness, we consider a stepwise distribution of $\beta(\theta)$ parameters (with the $\Rb$ dependence now absorbed in the paraxial angle $\theta$, see Appendix), which could be achieved by projecting a correspondingly shaped laser beam on an electron-transparent film at an oblique angle \cite{paper311} or, alternatively, by a weakly focused laser beam illuminating a film featuring a stepwise thickness profile consisting of concentric circular zones [see Fig.~\ref{Fig3}(a)]. This configuration reduces Eq.~(\ref{malpha3}) to Eq.~(\ref{malpha2}), with coefficients $a_i$ determined by restricting the $\theta$ integral to each of the $i$ zones.

We numerically optimize temporal compression at the focal spot $\rb=0$ (where the coefficients $a_i$ become real) by finding the maximum of ${\rm DOC}_1(\rb=0)$ \cite{notecompression1} through the steepest-gradient method. Separating the circle defined by the NA in the interaction plane into a total of $N$ equal-area zones [Fig.~\ref{Fig2}(a)], the coefficients $a_i$ are made independent of $i$. Under these conditions, the optimum focal electron wave function, represented over an optical period at $\rb=0$ in Fig.~\ref{Fig2}(b), becomes increasingly sharper as $N$ is increased (see Table~\ref{TableS2} in the Appendix for a subset of the obtained optimum values of $\beta_i$). Again, attainable values of the coupling coefficients $\beta_i$ are obtained. Interestingly, optimum results are obtained for a propagation distance $z=z_T/4$ (e.g., $z\approx2.5\,$mm for 60\,keV electrons and 4\,eV photons), which renders odd $\ell$ terms in quadrature relative to even terms. The corresponding degree of coherence is calculated analytically for each set of $\beta_i$ values (see Appendix) and plotted as a function of the number of zones $N$ and the harmonic order $m$ in Fig.~\ref{Fig2}(c). A monotonic increase is observed for each order $m$ as $N$ increases, and in particular, we find ${\rm DOC}_m>0.8$ for $|m|\le5$ with $N\ge15$.

The optimum solution at $\rb=0$ is compatible with spatial focusing, as revealed by the analysis presented in Fig.~\ref{Fig3}. Remarkably, a good level of temporal compression is simultaneously obtained within a finite spatial region covering the focal spot under the configuration presented in Fig.~\ref{Fig3}(a). At the time of maximum electron density [Fig.~\ref{Fig3}(b)], the focal spot is laterally confined within a region $R\lesssim\lambda_e/2\,{\rm NA}$ (e.g., $\approx1\,${\AA} for 100\,keV electrons and NA=0.02) for $N=30$, whereas it extends over a longitudinal range $|z|\lesssim\lambda_e/({\rm NA})^2$ ($\approx10\,$nm). When examining the temporal profile of cross sections passing by the focal spot and oriented along the transverse [Fig.~\ref{Fig3}(c)] and longitudinal [Fig.~\ref{Fig3}(d)] directions, we observe an overall level of compression similar to the optimized behavior at the spot center. A similar preservation of spatial focusing is observed with other values of $N$, while this parameter is primarily affecting temporal compression. For illustration, we present focal spot movies [Fig.~\ref{Fig3}(e)] revealing the emergence of a sharp electron density distribution within an interval spanning 20\% of the optical period for $N=2$, while shorter focal splot durations are obtained with larger values of $N$.

% =========================================================
\section{Conclusion}

In summary, we predict the formation of subnanometer-attosecond spatiotemporal electron probes by molding the transverse electron wave functions through PINEM-like interactions with spatially separated optical fields structured in relatively simple zone profiles. This approach is generally compatible with spatial electron focusing in scanning transmission electron microscopy, where the required optical fields could be directly projected on an electron-transparent plate. Alternatively, a simpler configuration could rely on illumination by a broad light beam, supplemented by lateral structuring of the plate (e.g., a dielectric film coated with metal and forming a layer of laterally varying thickness; see Appendix). While we have considered monochromatic light, such that a long electron pulse is transformed into a train of attosecond pulses spaced by the optical period, more general illumination conditions relying on broadband fields could be employed to obtain individual electron pulses with much wider temporal separation. As an extrapolation of these ideas, we envision the formation of arbitrary spatiotemporal electron profiles consisting, for instance, of several individual probes at tunable positions and instants to realize subnanometer-attosecond electron-electron pump-probe spectroscopy. A currently attainable light-electron pump-probe scheme could consist in triggering strongly nonlinear processes in a sample through ultrafast laser pulse irradiation, whose fast evolution within a sub-optical-cycle timescale could be probed by compressed electrons such as those here investigated.

\section*{ACKNOWLEDGMENTS}

We thank V. Di Giulio, A. Feist, J. H. Gaida, and S. V. Yalunin for insightful discussions. This work has been supported in part by the European Research Council (Advanced Grant 789104-eNANO), the European Commission (Horizon 2020 Grants FET-Proactive 101017720-EBEAM and FET-Open 964591-SMART-electron), the Spanish MICINN (PID2020-112625GB-I00 and Severo Ochoa CEX2019-000910-S), the Catalan CERCA Program, and Fundaci\'os Cellex and Mir-Puig, and the Humboldt Foundation.

\section*{APPENDIX}
%\appendix
\renewcommand{\thefigure}{A\arabic{figure}} %---SI
\renewcommand{\theequation}{A\arabic{equation}} %---SI
\renewcommand{\thetable}{A\arabic{table}} %---SI
\begin{widetext}

% =========================================================
% --- collimated e-beam interaction with lossy sample -----
% =========================================================
%\section{...} \label{sec1}

% ---------------------------------------------------------
\subsection{Paraxial free-electron focusing}

We consider a monochromated free-electron beam (e-beam) of kinetic energy $E_0$ moving along the positive $z$ direction. Starting with the electron wave function $\psi(\Rb',z_0)$ at a plane $z=z_0$ [Fig.~\ref{FigS1}(a)], where $\Rb'=(x',y')$ denotes the transverse position coordinates, we can construct the wave function $\psi(\Rb,z_1)$ at a different plane $z=z_1$ after free propagation by first projecting on components of transverse wave vector $\Qb=(q_x,q_y)$ and then incorporating the $z$ dependence through a longitudinal wave vector $q_z=\sqrt{q_0^2-Q^2}$, where $q_0=\me v\gamma/\hbar$ is the total electron wave vector corresponding to the specified kinetic energy $E_0=\me c^2(\gamma-1)$, written in terms of the velocity $v=c\sqrt{1-1/(E_0/\me c^2+1)^2}$ and the Lorentz factor $\gamma=1/\sqrt{1-v^2/c^2}$. More precisely,
\begin{align}
\psi(\Rb,z_1)&=\int \frac{d^2\Qb}{(2\pi)^2}\,\ee^{\ii\Qb\cdot\Rb+\ii q_z(z_1-z_0)}\int d^2\Rb' \ee^{-\ii\Qb\cdot\Rb'} \psi(\Rb',z_0) \nonumber\\
% &\approx\ee^{\ii q_0(z_1-z_0)}\int \frac{d^2\Qb}{(2\pi)^2}\,\ee^{\ii\Qb\cdot\Rb-\ii Q^2(z_1-z_0)/2q_0}\int d^2\Rb' \ee^{-\ii\Qb\cdot\Rb'} \psi(\Rb',z_0) \nonumber\\
&\approx\frac{-\ii q_0}{2\pi\,(z_1-z_0)}\ee^{\ii q_0(z_1-z_0)}\int d^2\Rb' \ee^{\ii q_0|\Rb-\Rb'|^2/2(z_1-z_0)} \psi(\Rb',z_0), \label{eq1}
\end{align}
where the second line is obtained by analytically integrating over $\Qb$ in the paraxial approximation $q_z\approx q_0-Q^2/2q_0$.

% Figure S1 -----------------------------------------------
\begin{figure*}[h] \label{FigS1}
\begin{centering} \includegraphics[width=1.0\textwidth]{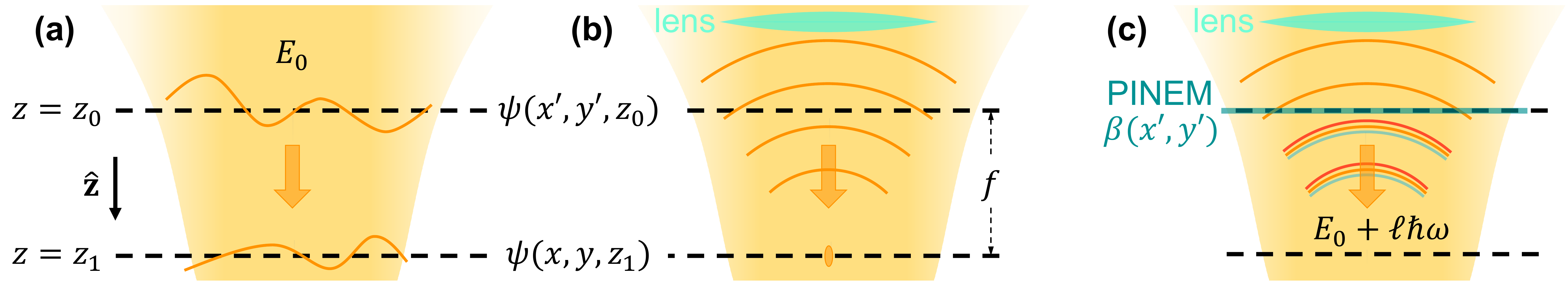} \par\end{centering}
\caption{{\bf Free-space electron propagation.} (a)~We consider free-space propagation of a monochromatic electron of kinetic energy $E_0$ from a plane $z=z_0$ to a plane $z=z_1$, where the wave function is $\psi(x',y',z_0)$ and $\psi(x,y,z_1)$, respectively. The e-beam is assumed to be paraxial and moving toward the positive $z$ direction. (b)~We are interested in the propagation of a beam focused at the $z=z_1$ plane through a lens placed before $z=z_0$. (c)~PINEM interaction is incorporated in the $z=z_0$ plane through a light-electron coupling coefficient $\beta(x',y')$ that depends on the transverse coordinates and produces a coherent electron comb of energies $E_0\pm\ell\hbar\omega$, separated from the incident one by multiples of the photon energy $\hbar\omega$.}
\end{figure*}

We are interested in an e-beam prepared to be focused at $\rb=0$ by an axially symmetric, aberration-free lens placed before the $z=z_0$ plane [Fig.~\ref{FigS1}(b)], so we write the incident wave function as $\psi(\Rb',z_0)=\psi_0\ee^{-\ii q_0 R'^2/2f}$, where $\psi_0$ is an overall amplitude coefficient and $f$ is the focal distance relative to such plane. Introducing these elements in Eq.~(\ref{eq1}), the electron wave function near the focal region reduces to
\begin{align}
\psi(\rb)=\frac{-\ii\psi_0q_0}{f+z}
\,\ee^{\ii (q_0R^2/2)/(f+z)}
\,\ee^{\ii q_0(f+z)}
\int_0^\Rmax\!\!\! R'dR'
\,J_0\big[q_0RR'/(f+z)\big]
\,\ee^{-\ii(q_0zR'^2/2f)/(f+z)}, \label{eq2}
\end{align}
where the longitudinal coordinate $z=z_1-z_0-f$ is referred to the focal plane and the upper limit of the radial integral $\Rmax\approx NA\times f$ is limited by the numerical aperture (NA$\lesssim0.02$ in typical transmission electron microscopes).

%\newcommand*{\widebox}[2][0.5em]{\fbox{\hspace{#1}$\displaystyle #2$\hspace{#1}}}
%\begin{align}[box=\widebox] ... \nonumber \end{align}

% ---------------------------------------------------------

% ---------------------------------------------------------
\subsection{Time-dependent focused beam after coherent PINEM interaction}

We now introduce an $\Rb'$-dependent PINEM interaction with light of frequency $\omega$ at the $z=z_0$ plane [Fig.~\ref{FigS1}(c)], expressed through a coupling coefficient $\beta(\Rb')$ \cite{paper371}. It is then convenient to explicitly write the time-dependent wave function, which, for the incident monochromatic electron, reads $\psi^{\rm inc}(\rb)\ee^{-\ii E_0t/\hbar}$. Right after the noted interaction, the wave function at $z=z_0$ becomes
\begin{align}
\psi(\Rb',z_0,t)=\psi^{\rm inc}(\Rb',z_0)\sum_{\ell=-\infty}^\infty J_\ell\big[2|\beta(\Rb')|\big]\,\ee^{\ii\ell{\rm arg}\{-\beta(\Rb')\}}\,\ee^{-\ii(E_0+\ell\hbar\omega)t/\hbar}. \label{eq3}
\end{align}
Each $\ell$ component in Eq.~(\ref{eq3}) needs to be propagated to the focal region according to Eq.~(\ref{eq1}), but with the electron wave vector $q_0$ replaced by $q_\ell=q_0+\ell\omega/v-2\pi\ell^2/z_T+\cdots$, corresponding to the modified electron energy $E_0+\ell\hbar\omega$, where $z_T=4\pi\me v^3\gamma^3/\hbar\omega^2$ is the so-called Talbot distance \cite{paper360}. Assuming axial symmetry in $\beta(\Rb')\equiv\beta(R')$ and applying Eq.~(\ref{eq2}) to propagate the wave function in Eq.~(\ref{eq3}) [i.e., considering again an incident focused beam characterized by a wave function $\psi^{\rm inc}(\Rb',z_0)=\psi_0\ee^{-\ii q_0 R'^2/2f}$], the time-dependent electron wave function in the focal region reads
\begin{align}
\psi(\rb,t)&=\frac{-\ii\psi_0}{f+z}
\sum_\ell q_\ell
\,\ee^{\ii (q_\ell R^2/2)/(f+z)}
\,\ee^{\ii q_\ell(f+z)-\ii(E_0/\hbar+\ell\omega)t} \label{fullpsi}\\
&\quad\quad\quad\times\int_0^\Rmax \!\!\! R'dR'
\,J_0\big[q_\ell RR'/(f+z)\big]
\,\ee^{-\ii(q_\ell zR'^2/2f)/(f+z)}
\,J_\ell\big[2|\beta(R')|\big]
\,\ee^{\ii\ell{\rm arg}\{-\beta(R')\}}. \nonumber
\end{align}
This expression can be simplified by assuming parameter ranges that encompass a broad set of experimental conditions, such as $E_0\ge1\,$keV, NA$\lesssim0.02$, $f\gtrsim1\,$mm, and a photon wavelength $\lambda_0=2\pi c/\omega\sim1\,\mu$m, for which the electron wavelength is $\lambda_e<39\,$pm$\,\ll\lambda_0$, the Talbot distance is $z_T\gtrsim0.2\,$mm, and the radial and longitudinal extensions of the focal spot are delimited by $R\lesssim\lambda_e/{\rm NA}$ and $|z|\lesssim\lambda_e/({\rm NA})^2$. Neglecting phase contributions of the order of $\lambda_e/\lambda_0$, $z/f$, and $R/R_{\rm max}$ (see detailed analysis in Table~\ref{TableS1}), we find
\begin{subequations}
\label{psialphar}
\begin{align}
&\psi(\rb,t) \approx C\sum_\ell \alpha_\ell(\rb)
\,\ee^{\ii\ell\omega(f+z-vt)/v}, \label{psir}\\
&\alpha_\ell(\rb)=\ee^{-\ii\,\ell^2\phi}\int_0^{\rm NA} \!\!\! \theta d\theta \,J_0(q_0R\,\theta) \,\ee^{-\ii q_0z\,\theta^2/2} \,J_\ell\big[2|\beta(\theta)|\big] \,\ee^{\ii\ell{\rm arg}\{-\beta(\theta)\}}, \label{alphar}
\end{align}
\end{subequations}
where $\phi=2\pi\,f/z_T$ is a propagation phase accounting for velocity dispersion in different $\ell$ contributions, we have changed the variable of integration to $\theta=R'/f$ (limited by ${\rm NA}\approx\Rmax/f$), the $R'$ dependence of the PINEM coefficient is indicated through $\beta(\theta)$, and the coefficient $C=-\ii\,\psi_0q_0f\,\ee^{\ii q_0(f+z+R^2/2f)}\ee^{-\ii(E_0/\hbar)t}$ has a constant modulus (independent of $\rb$ and $t$).

% Table S1 ------------------------------------------------
\begin{table*}[t]
\begin{tabular}{lll}
\hline \\
Electron & \quad\quad &
$q_\ell=q_0[1+\ell\eta-\ell^2\eta'+\cdots]$
{\red---} $\eta=\omega/vq_0=(c/v)\lambda_e/\lambda_0\ll1$
{\red---} $\eta'=2\pi/q_0z_T$
\\ momentum & & {\red---} $\eta'/\eta=2\pi v/\omega z_T =(v/c)\lambda_0/z_T =(c/2v\gamma^2)(\lambda_e/\lambda_0)\ll1$
\\ expansion & & {\red---} $z_T=4\pi\me v^3\gamma^3/\hbar\omega^2=2(v\gamma/c)^2(\lambda_0^2/\lambda_e)$
{\red---} $q_0=\me v\gamma/\hbar=2\pi/\lambda_e$ \\ \\ \hline \\
Focal region & &
$R\lesssim g \lambda_e/({\rm NA})$
{\red---} $|z|\lesssim g' \lambda_e/({\rm NA})^2$
{\red---} ${\rm NA}\approx\Rmax/f$
{\red---} $g,g'\sim1$
\\ \\ \hline \\
Approximations & &
${\red (q_\ell R^2/2)/(f+z)}=(q_0R^2/2f)[1+\ell\eta-\ell^2\eta'+\cdots][1-z/f+z^2/f^2+\cdots]\approx q_0R^2/2f$
   \\& & $+O\big[\eta q_0R^2/2f=\pi(c/v)R^2/f\lambda_0\lesssim\pi g(c/v)(R/\Rmax)(\lambda_e/\lambda_0)\ll1\big]$
     {\gray {\red---} $R/f\ll R'/f\le{\rm NA}\ll1$}
   \\& & $+O\big[q_0|z|R^2/2f^2=\pi|z|R^2/f^2\lambda_e\lesssim\pi g(|z|/f)(R/\Rmax)\ll1\big]$
     {\gray {\red---} $|z|/f\ll1$}
     {\gray {\red---} $R/\lambda_0\ll1$} \\
 & &
${\red q_\ell RR'/(f+z)}=(q_0RR'/f)[1+\ell\eta-\ell^2\eta'+\cdots][1-z/f+z^2/f^2+\cdots]\approx q_0RR'/f$
   \\& & $+O\big[\eta q_0RR'/f=2\pi(c/v)RR'/f\lambda_0<2\pi g(c/v)\,\lambda_e/\lambda_0\ll1\big]$
   \\& & $+O\big[q_0|z|RR'/f^2=2\pi|z|RR'/f^2\lambda_e\lesssim2\pi g(|z|/f)(R'/\Rmax)\le2\pi g(|z|/f)\ll1\big]$ \\
 & &
${\red (q_\ell zR'^2/2f)/(f+z)}=(q_0zR'^2/2f^2)[1+\ell\eta-\ell^2\eta'+\cdots][1-z/f+z^2/f^2+\cdots]\approx q_0zR'^2/2f^2$
   \\& & $+O\big[\eta q_0|z|R'^2/2f^2=\pi(c/v)|z|R'^2/f^2\lambda_0\le\pi(c/v)|z|({\rm NA})^2/\lambda_0\lesssim\pi g'(c/v)\lambda_e/\lambda_0\ll1\big]$
   \\& & $+O\big[q_0z^2R'^2/2f^3\le q_0z^2({\rm NA})^2/2f\lesssim g'q_0\lambda_e|z|/2f=\pi g'|z|/f\ll1\big]$ \\
 & &
${\red q_\ell(f+z)}=q_0(f+z)(1+\ell\eta-\ell^2\eta'+\cdots)=q_0(f+z)+q_0(f+z)\ell\eta-q_0(f+z)\ell^2\eta'+\cdots$
   \\& & $\approx q_0(f+z)+(f+z)\ell\omega/v-2\pi f\ell^2/z_T+\cdots$
     {\gray {\red---} $q_0|z|\eta'=2\pi|z|/z_T<\pi g'(c/v\gamma)^2\,\lambda_e^2/(\lambda_0\times{\rm NA})^2\ll1$} \\ \\ \hline
\end{tabular}
\caption{Details of the approximations made to transform Eq.~(\ref{fullpsi}) into Eqs.~(\ref{psialphar}).}
\label{TableS1}
\end{table*}

The wave function in Eqs.\ (\ref{psialphar}) is periodic in time with a period $\tau=2\pi/\omega$ determined by the light frequency. The time-averaged probability density then reduces to
\begin{align}
\frac{1}{\tau}\int_0^\tau dt\,|\psi(\rb,t)|^2=|C|^2\sum_\ell |\alpha_\ell(\rb)|^2. \nonumber
\end{align}
In addition, upon integration over transverse coordinates, this quantity yields an electron current proportional to
\begin{align}
\frac{1}{\tau}\int d^2\Rb\int_0^\tau dt\,|\psi(\rb,t)|^2=\pi|\psi_0|^2R_{\rm max}^2, \label{Rintegratedpsi2}
\end{align}
where we have used the equation $\int_0^\infty xdx\,J_\ell(ax)J_\ell(a'x)=\delta(a-a')/a$ to reduce the double sum over sidebands to a single one, and then applied the identity $\sum_\ell J_\ell^2(a)=1$. Reassuringly, the result in Eq.~(\ref{Rintegratedpsi2}) is independent of the longitudinal position $z$, as expected from the conservation of electron probability.

It is convenient to discretize the $\theta$ dependence of $\beta(\theta)$ by considering $N$ concentric circular zones, such that $\beta(\theta)=\beta_i$ is uniform within each zone $\theta_{i-1}<\theta<\theta_i$, with $i=1,\cdots,N$ and $\theta_N={\rm NA}$. We also define $\theta_0=0$ to refer to the intersection with the axis of rotational symmetry. This allows us to rewrite Eq.~(\ref{alphar}) as
\begin{subequations}
\label{psialphai}
\begin{align}
&\alpha_\ell(\rb)=\ee^{-\ii\,\ell^2\phi}\sum_i a_i \,J_\ell(2|\beta_i|) \,\ee^{\ii\ell{\rm arg}\{-\beta_i\}}, \label{psii}\\
&a_i=\int_{\theta_{i-1}}^{\theta_i} \!\!\! \theta d\theta \,J_0(q_0R\,\theta) \,\ee^{-\ii q_0z\,\theta^2/2}. \label{alphai}
\end{align}
\end{subequations}
In particular, at the focal point $\rb=0$ the expansion coefficients are $a_i=(\theta_i^2-\theta_{i-1}^2)/2$, real and proportional to the areas of the concentric zones. Equations~(\ref{psialphai}) provide a simple prescription to parametrize any arbitrary $\beta(\theta)$ profile by taking a sufficiently large number of zones $N$. In the present work, we consider moderate values of $N$ under the assumption that the coupling coefficient is made uniform in each zone.

% ---------------------------------------------------------
\subsection{Evaluation of the degree of coherence}

We consider an electron modulated as shown in Eqs.~(\ref{psialphar}). When one is interested in the subsequent electron interaction with a specimen, the coherence factor that is defined as \cite{paper374,paper371}
\begin{align}
M_m(\rb)=\int_0^\tau dt |\psi(\rb,t)|^2\,\ee^{\ii m\omega t} \nonumber
\end{align}
provides a measure of its ability to excite an optical mode of frequency $m\omega$ (a harmonic $m$ of the light frequency) localized at a position $\rb$. In the process carried out in the main text to optimize the temporal compression of the electron, we maximize the degree of coherence \cite{paper374}
\begin{align}
{\rm DOC}_m(\rb)=|M_m(\rb)/M_0(\rb)|^2, \label{DOC}
\end{align}
which determines the enhancement in the excitation probability relative to an unmodulated electron. In the limit of a point particle, we have ${\rm DOC}_m(\rb)=1$ for all $m$'s. In practice, when maximizing ${\rm DOC}_1(0)$ (i.e., for $m=1$ at the focal spot), we obtain compressed electron pulses in which ${\rm DOC}_m(\rb)$ is also enhanced for other values of $m$ within an extended focal region, as illustrated by Figs.~\ref{Fig2} and \ref{Fig3} in the main text.

Considering light-coupling coefficients structured in a set of concentric circular zones, we start from Eq.~(\ref{psir}) to write the coherence factor as
\begin{align}
M_m(\rb)=|C|^2\sum_\ell \alpha_\ell(\rb)\alpha_{\ell+m}^*(\rb). \nonumber
\end{align}
A useful expression can be found by expanding $\alpha_\ell(\rb)$ as shown in Eq.~(\ref{psii}) and then making use of Graf's theorem (see Eq.~(9.1.79) of Ref.~\cite{AS1972}) to evaluate the $\ell$ sum. This leads to
\begin{align}
M_m(\rb)=|C|^2\sum_{ii'} a_ia_{i'}^*\,\ee^{\ii m(m\phi-\varphi_{i'}+\chi_{ii'})}J_m(2b_{ii'}), \label{Mm}
\end{align}
where $\varphi_i={\rm arg}\{-\beta_i\}$, $b_{ii'}=\sqrt{|\beta_i|^2+|\beta_{i'}|^2-2|\beta_i\beta_{i'}|\cos\xi_{ii'}}$ with $\xi_{ii'}=\varphi_i-\varphi_{i'}+2m\phi$, and we define the phase $\chi_{ii'}$ to satisfy the equations $b_{ii'}\sin\chi_{ii'}=|\beta_i|\sin\xi_{ii'}$ and $b_{ii'}\cos\chi_{ii'}=|\beta_{i'}|-|\beta_i|\cos\xi_{ii'}$. Under uniform illumination (i.e., $\beta_i$ independent of $i$), direct inspection of Eq.~(\ref{Mm}) leads to the single-PINEM result \cite{paper371} $M_m=(\sum_i|a_i|^2)|C|^2\ii^m\ee^{-\ii m\,{\rm arg}\{-\beta_1\}}J_m[4|\beta_1|\sin(m\phi)]$, for which the maximum degree of coherence is ${\rm DOC}_m=J_m^2(\zeta_m)$, obtained with $|\beta_1|=\zeta_m/|4\sin(m\phi)|$, where $\zeta_m$ is the maximum of $J_m(x)$ (e.g., $\zeta_1\approx1.8412$ for $m=1$, which leads to ${\rm DOC}_1\approx0.3386$).

% Similar to the single-PINEM configuration, the propagation phase $\phi=2\pi f/z_T$ plays an important role in the optimization of temporal compression through parallel-PINEM interactions. 

As an interesting configuration, we consider two concentric zones ($N=2$) with the central one having $\beta_1=0$ (i.e., no interaction with light), for which Eqs.~(\ref{DOC}) and (\ref{Mm}) produce
\begin{align}
{\rm DOC}_1=\Delta^2\frac{2J_1(2|\beta_2|)\sin\phi+\Delta J_1(4|\beta_2|\sin\phi)}{1+\Delta^2+2\Delta J_0(2|\beta_2|)}, \nonumber
\end{align}
where $\Delta=a_2/a_1$ is taken to be real. This expression has an absolute maximum of ${\rm DOC}_1\approx0.513$ for $\Delta\approx2.303$, $|\beta_2|\approx1.694$, and $\phi\approx0.284$, already exceeding the single-PINEM result.

% Table S2 ------------------------------------------------
\begin{table*}[t]
\tiny
\begin{tabular}{l |c r |c r |c r |c r |c r |c r |c r |c r |c r}
{\scriptsize $N$} & {\scriptsize $|\beta_{1}|$} & {\scriptsize ${\rm arg}\{\beta_{1}\}$} & {\scriptsize $|\beta_{2}|$} & {\scriptsize ${\rm arg}\{\beta_{2}\}$} & {\scriptsize $|\beta_{3}|$} & {\scriptsize ${\rm arg}\{\beta_{3}\}$} & {\scriptsize $|\beta_{4}|$} & {\scriptsize ${\rm arg}\{\beta_{4}\}$} & {\scriptsize $|\beta_{5}|$} & {\scriptsize ${\rm arg}\{\beta_{5}\}$} & {\scriptsize $|\beta_{6}|$} & {\scriptsize ${\rm arg}\{\beta_{6}\}$} & {\scriptsize $|\beta_{7}|$} & {\scriptsize ${\rm arg}\{\beta_{7}\}$} & {\scriptsize $|\beta_{8}|$} & {\scriptsize ${\rm arg}\{\beta_{8}\}$} & {\scriptsize $|\beta_{9}|$} & {\scriptsize ${\rm arg}\{\beta_{9}\}$} \\ \hline
{\bf 1} & {\bf 0.460} & {\bf   0.0$^{\circ}$} & - & -\;\;\;\;\; & - & -\;\;\;\;\; & - & -\;\;\;\;\; & - & -\;\;\;\;\; & - & -\;\;\;\;\; & - & -\;\;\;\;\; & - & -\;\;\;\;\; & - & -\;\;\;\;\; \\
{\bf 2} & {\bf 0.947} & {\bf   0.0$^{\circ}$} & {\bf 0.947} & {\bf 128.0$^{\circ}$} & - & -\;\;\;\;\; & - & -\;\;\;\;\; & - & -\;\;\;\;\; & - & -\;\;\;\;\; & - & -\;\;\;\;\; & - & -\;\;\;\;\; & - & -\;\;\;\;\; \\
{\bf 3} & {\bf 1.204} & {\bf   0.0$^{\circ}$} & {\bf 0.484} & {\bf  72.4$^{\circ}$} & {\bf 1.204} & {\bf 144.8$^{\circ}$} & - & -\;\;\;\;\; & - & -\;\;\;\;\; & - & -\;\;\;\;\; & - & -\;\;\;\;\; & - & -\;\;\;\;\; & - & -\;\;\;\;\; \\
{\bf 4} & {\bf 0.540} & {\bf   0.0$^{\circ}$} & {\bf 2.039} & {\bf  37.2$^{\circ}$} & {\bf 2.039} & {\bf 239.7$^{\circ}$} & {\bf 0.540} & {\bf 276.9$^{\circ}$} & - & -\;\;\;\;\; & - & -\;\;\;\;\; & - & -\;\;\;\;\; & - & -\;\;\;\;\; & - & -\;\;\;\;\; \\
{\bf 5} & {\bf 2.248} & {\bf   0.0$^{\circ}$} & {\bf 2.248} & {\bf 161.1$^{\circ}$} & {\bf 0.975} & {\bf 140.4$^{\circ}$} & {\bf 0.975} & {\bf  20.8$^{\circ}$} & {\bf 0.273} & {\bf  80.6$^{\circ}$} & - & -\;\;\;\;\; & - & -\;\;\;\;\; & - & -\;\;\;\;\; & - & -\;\;\;\;\; \\
{\bf 6} & {\bf 1.114} & {\bf   0.0$^{\circ}$} & {\bf 1.114} & {\bf 139.4$^{\circ}$} & {\bf 1.171} & {\bf 139.2$^{\circ}$} & {\bf 3.162} & {\bf 346.9$^{\circ}$} & {\bf 3.162} & {\bf 152.5$^{\circ}$} & {\bf 1.171} & {\bf   0.2$^{\circ}$} & - & -\;\;\;\;\; & - & -\;\;\;\;\; & - & -\;\;\;\;\; \\
{\bf 7} & {\bf 3.330} & {\bf   0.0$^{\circ}$} & {\bf 1.703} & {\bf 157.3$^{\circ}$} & {\bf 1.108} & {\bf   8.1$^{\circ}$} & {\bf 1.108} & {\bf 158.6$^{\circ}$} & {\bf 0.496} & {\bf  83.4$^{\circ}$} & {\bf 3.330} & {\bf 166.7$^{\circ}$} & {\bf 1.703} & {\bf   9.5$^{\circ}$} & - & -\;\;\;\;\; & - & -\;\;\;\;\; \\
{\bf 8} & {\bf 3.620} & {\bf   0.0$^{\circ}$} & {\bf 0.423} & {\bf  84.0$^{\circ}$} & {\bf 3.620} & {\bf 168.0$^{\circ}$} & {\bf 0.423} & {\bf  84.0$^{\circ}$} & {\bf 2.030} & {\bf 159.5$^{\circ}$} & {\bf 1.575} & {\bf   4.0$^{\circ}$} & {\bf 1.575} & {\bf 164.0$^{\circ}$} & {\bf 2.030} & {\bf   8.5$^{\circ}$} & - & -\;\;\;\;\; \\
{\bf 9} & {\bf 2.440} & {\bf   0.0$^{\circ}$} & {\bf 0.404} & {\bf  80.7$^{\circ}$} & {\bf 0.753} & {\bf 139.4$^{\circ}$} & {\bf 4.520} & {\bf 355.7$^{\circ}$} & {\bf 0.753} & {\bf  21.9$^{\circ}$} & {\bf 2.440} & {\bf 161.3$^{\circ}$} & {\bf 2.440} & {\bf   0.0$^{\circ}$} & {\bf 4.520} & {\bf 165.6$^{\circ}$} & {\bf 2.440} & {\bf 161.3$^{\circ}$} \\
\end{tabular}
\caption{\textnormal{Optimum values of the $\beta_i$ parameters corresponding to the solutions presented in Fig.~\ref{Fig2} of the main text for $N$ concentric circular zones of equal area with $N=1-9$. The optimum dispersive phase is $\phi=2\pi z/z_T=\pi/4$ in all cases.}}
\label{TableS2}
\end{table*}

% ---------------------------------------------------------
\subsection{Optimum light-electron coupling parameters for concentric circular zones of equal area}

We use the steepest-gradient method to optimize the degree of coherence for $m=1$ at the focal point $\rb=0$ [i.e., ${\rm DOC}_1(0)$] as an approach to obtain temporally compressed electron wave function profiles. In particular, we consider configurations consisting of a number $N$ of concentric zones with the same area, such that the coefficients $a_i$ cancel out in the evaluation of Eq.~(\ref{DOC}) with $M_m(0)$ calculated from Eq.~(\ref{Mm}). More precisely, we find
\begin{align}
{\rm DOC}_1(0)=\left|\frac{\sum_{ii'}\ee^{\ii(\phi-\varphi_{i'}+\chi_{ii'})}J_1(2b_{ii'})}{\sum_{ii'}J_0(2b_{ii'})}\right|^2. \nonumber
\end{align}
In all cases, we obtain an optimum dispersive phase $\phi=2\pi z/z_T=\pi/4$. The so-obtained optimum light-electron coupling parameters are listed in Table~\ref{TableS2} for $N=1-9$.

% Figure S2 -----------------------------------------------
\begin{figure*}[h]
\begin{centering} \includegraphics[width=0.85\textwidth]{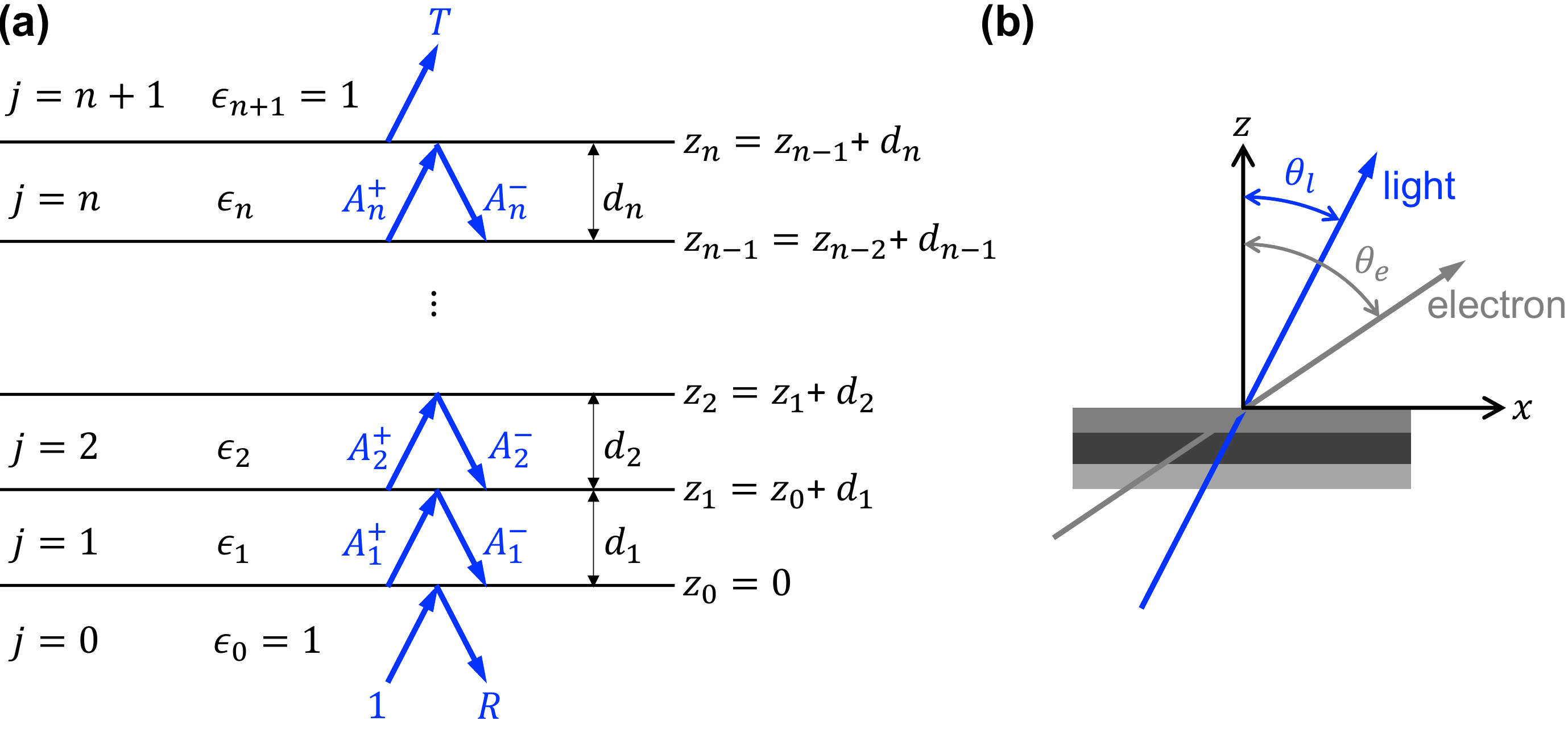} \par\end{centering}
\caption{{\bf Light-electron coupling in a planar multilayer structure.} (a)~Parameters defining the multilayer structure and the propagation of light throughout it. (b)~Orientation of the incidence electron and light directions relative to the multilayer.}
\label{FigS2}
\end{figure*}

% ---------------------------------------------------------
\subsection{Light-electron coupling coefficient in planar multilayers under plane-wave illumination}

A practical realization of $i=1,\cdots,N$ zones featuring different light-electron coupling coefficients $\beta_i$ could be based on illumination by a uniform light plane wave of amplitude $\mathcal{E}_0$, such that each zone $i$ consists of a planar multilayer with thicknesses adjusted to obtain the desired coefficient $\beta_i$ up to a global factor $\mathcal{E}_0$. In this section, we sketch a calculation of the light-electron coupling coefficient mediated by a planar multilayer under oblique light and electron incidence conditions. We set the surface normal along $z$ and, for simplicity, take the light and electron incidence directions in the $x$-$z$ plane, forming angles $\theta_l$ and $\theta_e$ relative to the $z$ axis, respectively, with the light incident from the $z<0$ region [see Fig.~\ref{FigS2}(b)].

We consider $n$ layers ($j=1,\cdots,n$) of thicknesses $d_j$ and permittivities $\epsilon_j$, as shown in Fig.~\ref{FigS2}(a). The optical electric field in each layer $j$ delimited by $z_{j-1}<z<z_j$ is expressed in terms of plane wave coefficients $A_j^\pm$ by writing $\Eb(\rb,t)=2{\rm Re}\big\{\Eb(\rb)\,\ee^{-\ii\omega t}\big\}$ with
\begin{subequations}
\label{fields}
\begin{align}
\Eb(\rb)=\mathcal{E}_0\,\left[A_j^+\,\eh^+_j\,\ee^{\ii k_{jz}(z-z_{j-1})}+A_j^-\,\eh^-_j\,\ee^{-\ii k_{jz}(z-z_j)}\right]\;\ee^{\ii k_xx}, \quad\quad\quad z_{j-1}<z<z_j
\end{align}
where the $x$ component of the wave vector $k_x>0$ and the frequency $\omega$ are conserved during light propagation, $k_{jz}=\sqrt{k_j^2-k_x^2+\ii0^+}$ with $k_j=(\omega/c)\sqrt{k_j}$ and ${\rm Im}\{k_{jz}\}>0$ is the out-of-plane wave vector, and $\eh^\pm_j=(\pm k_{jz}\xx-k_x\zz)/k_j$ are upward ($+$ sign) and downward ($-$ sign) vectors for p polarization inside medium $j$. We dismiss s-polarized fields because they produce a vanishing light-electron coupling under the geometry of Fig.~\ref{FigS2}(b). In the near-side region ($z<0$, $j=0$, $\epsilon_0=1$), we write the field as
\begin{align}
\Eb(\rb)=\mathcal{E}_0\,\left(\eh^+_0\,\ee^{\ii k_zz}+R\;\eh^-_0\,\ee^{-\ii k_zz}\right)\;\ee^{\ii k_x x},  \quad\quad\quad z<0
\end{align}
where $R$ is the reflection coefficient of the entire multilayer and $k_z\equiv k_{0z}$. Likewise, the field in the far-side region ($z>z_n$, $j=n+1$, $\epsilon_{n+1}=1$) reads
\begin{align}
\Eb(\rb)=\mathcal{E}_0\,T\;\eh^+_0\,\ee^{\ii[k_z(z-z_n)+k_xx]},  \quad\quad\quad z>z_n
\end{align}
\end{subequations}
where $k_z$ and $\eh^+_0=\cos\theta_l\,\xx-\sin\theta_l\,\zz$ are the same as in the near side and $T$ is the transmission coefficient.

The coefficients $A_j^\pm$, $R$, and $T$ in Eqs.~(\ref{fields}) are determined by the boundary conditions at the $z=z_j$ interfaces. Here, we consider the equivalent Fabry-Perot-like expressions
\begin{align}
A_j^+=r_{j,j-1}\,\ee^{\ii k_{jz}d_j}\,A_j^-+t_{j-1,j}\,\ee^{\ii k_{{j-1},z}d_{j-1}}\,A_{j-1}^+, \nonumber\\
A_j^-=r_{j,j+1}\,\ee^{\ii k_{j,z}d_j}\,A_j^++t_{j+1,j}\,\ee^{\ii k_{{j+1},z}d_{j+1}}\,A_{j+1}^- \nonumber
\end{align}
for $j=1,\cdots,n$, written in terms of the p-polarization Fresnel reflection and transmission coefficients $r_{j,j'}=(\epsilon_{j'}k_{j,z}-\epsilon_{j}k_{j',z})/(\epsilon_{j'}k_{j,z}+\epsilon_{j}k_{j',z})$ and $t_{j,j'}=2\sqrt{\epsilon_{j}\epsilon_{j'}}\,k_{j,z}/(\epsilon_{j'}k_{j,z}+\epsilon_{j}k_{j',z})$, respectively, at each $j|j'$ interface for incidence from the $j$ side. We are left with a linear system of $2n$ equations and variables (the coefficients $A_j^\pm$ with $j=1,\cdots,n$), supplemented by $A_0^+=1$ (corresponding to an incident field amplitude $\mathcal{E}_0$) and $A_{n+1}^-=0$ (no incident wave from the far side), as well as the parameters $d_0=d_{n+1}=0$, which are defined such that the above equations take a compact form. After finding $A_j^\pm$ by using standard linear algebra techniques, the reflection and transmission coefficients are obtained from $R=r_{0,1}+t_{1,0}\,\ee^{\ii k_{1,z}d_1}\,A_1^-$ and $T=t_{n,n+1}\,\ee^{\ii k_{nz}d_n}\,A_n^+$ [see Fig.~\ref{FigS2}(a)].

The light-electron coupling coefficient is given by
\begin{align}
\beta=\frac{ev}{\hbar\omega}\int_{-\infty}^\infty dt\; \vv\cdot\Eb(\rb=\vb t)\,\ee^{-\ii\omega t},
\label{beta}
\end{align}
where $\vb=v\vv=v_x\,\xx+v_z\,\zz$ with $v_x=v\sin\theta_e$ and $v_z=v\cos\theta_e$ is the velocity vector oriented along the direction $\vv=\sin\theta_e\,\xx+\cos\theta_e\,\zz$, and we take the electron to cross the $z=0$ plane at $t=0$. This expression reduces to Eq.~(\ref{mbeta}) in the main text when $z$ is chosen along the e-beam direction. Finally, inserting Eqs.~(\ref{fields}) into Eq.~(\ref{beta}), we find
\begin{align}
\beta=\pm\frac{\ii ev\mathcal{E}_0}{\hbar\omega^2}\bigg\{
&\sum_{j=1}^n\bigg[
A_j^+\,\vv\cdot\eh^+_j\;
\frac{\ee^{\ii(k_xv_x-\omega)z_j/v_z}\ee^{\ii k_{jz}d_j}-\ee^{\ii(k_xv_x-\omega)z_{j-1}/v_z}}
{1-(k_xv_x+k_{jz}v_z)/\omega} \nonumber\\
&\;\;\;\;\,+A_j^-\,\vv\cdot\eh^-_j\;
\frac{\ee^{\ii(k_xv_x-\omega)z_j/v_z}-\ee^{\ii(k_xv_x-\omega)z_{j-1}/v_z}\ee^{\ii k_{jz}d_j}}
{1-(k_xv_x-k_{jz}v_z)/\omega}\bigg] \nonumber\\
-&\sin(\theta_l-\theta_e)\;\frac{1-T\,\ee^{\ii(k_xv_x-\omega)z_n/v_z}}{1-\cos(\theta_l-\theta_e)\,v/c} \nonumber\\
-&\sin(\theta_l+\theta_e)\;\frac{R}{1+\cos(\theta_l+\theta_e)\,v/c} \bigg\},
\label{betafinal}
\end{align}
where the overall $+$ and $-$ signs apply to electrons moving along upward ($-\pi/2<\theta_e<\pi/2$) or downward ($\pi/2<\theta_e<3\pi/2$) directions, respectively, while the light is propagating upwardly ($0<\theta_l<\pi/2$) in all cases.

% Figure S3 -----------------------------------------------
\begin{figure*}[h]
\begin{centering} \includegraphics[width=1.00\textwidth]{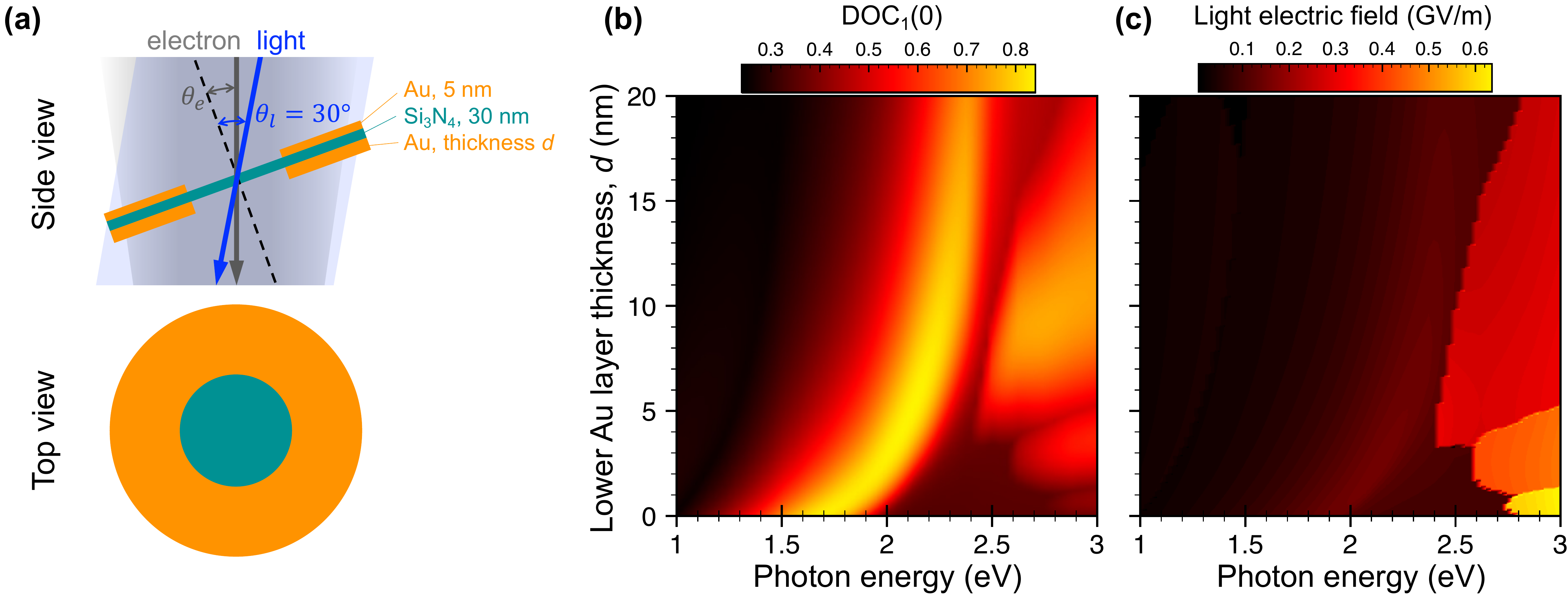} \par\end{centering}
\caption{{\bf Two-zone plate design.} (a)~We consider a structure consisting of a 20 nm Si$_3$N$_4$ film partially coated with Au layers on both sides: 5 nm on top and a layer of thickness $d$ below. The coated and uncoated regions are taken to have the same area. The plate is optimized for 200\,keV electrons, light incidence at an angle $\theta_l=30^\circ$ relative to the surface normal, and electrons incident with an angle $\theta_e=\arcsin\big[(v/c)\sin\theta_l\big]\approx20.3^\circ$ such that the light-imprinted and electron-dispersion-corrected phase has a vanishing overall geometric factor $\propto(\omega/c)\sin\theta_l-(\omega/v)\sin\theta_e=0$, independent of lateral position along the plate. (b)~We vary the photon energy $\hbar\omega$ and the lower Au layer thickness $d$ to explore the resulting degree of coherence DOC$_1(0)$ at the focal point under the conditions of Fig.~\ref{Fig2} in the main text. The obtained values are close to the absolute limit for two-zone plates over a wide range of thicknesses. (c)~The actual value of DOC$_1(0)$ depends on the light electric field amplitude, which we adjust for each combination of $\omega$ and $d$ to optimize temporal compression. The so-obtained values of the field amplitude are attainable with synchronized light-electron pulses in ultrafast electron microscopes. The plate-focus distance is taken to be $z_T/4=\pi\me v^3\gamma^3/\hbar\omega^2$ in all cases (e.g., $z_T/4\approx23\,$mm at $\hbar\omega=2\,$eV).}
\label{FigS3}
\end{figure*}

% ---------------------------------------------------------
\subsection{Two-zone design}

We now use the light-coupling coefficient given by Eq.~(\ref{betafinal}) for multilayer structures to optimize a specific plate yielding a reasonable level of temporal compression. For simplicity, we target a two-zone plate with the structure presented in Fig.~\ref{FigS3}(a). Plane-wave illumination requires oblique incidence to guarantee a nonvanishing light-electron coupling. In addition, electrons have to impinge obliquely as well to cancel the geometric optical phase. The phase-cancellation condition is $v\sin\theta_l=c\sin\theta_e$. We explore the performance of the plate with different thicknesses of one of the coating layers for a $1-3\,$eV photon energy range, yielding a maximum DOC$_1(0)$ close to the absolute limit for two-zone plates over a wide range of thicknesses and photon energies around 2\,eV. Similar results are obtained for different e-beam energies and geometrical parameters, which can be varied to move the spectral region showing an optimum degree of coherence.

\end{widetext}

%\bibliography{../../../bibtex/refsL.bib}


\end{document}